\documentclass{PoS}

\title{Spectral properties of a sample of type 1 AGNs: influence of star  
                             formation}

\ShortTitle{Spectral properties of a sample of type 1 AGN}

\author{\speaker{Luka \v C. Popovi\'c}, Jelena Kova\v cevi\'c and Milan S. Dimitrijevi\'c\thanks{The Fe II template can be found at: http://servo.aob.rs/FeII\_AGN/}\\
        Astronomical Observatory\\
        E-mail: \email{lpopovic@aob.rs}}

\abstract{To find the spectral properties of AGNs in optical spectral band (around the H$\beta$ line) we constructed an Fe II template, that is covering
Fe II  emission from 4100 \AA\ to 5600 \AA.
Using the new Fe II template we explored spectral properties of 302 type 1 AGNs. The most interesting results we found is the correlation of the H$\beta$ Full Width at Half Maximum (FWHM) and 
luminosity for a subsample of type 1 AGNs where the ratio of narrow lines indicates a significant starburst contribution.}

\FullConference{Workshop on: Nuclei of Seyfert galaxies and QSOs - Central engine \& conditions of star formation,\\
		November 6-8, 2012\\
		Max-Planck-Insitut für Radioastronomie (MPIfR), Bonn, Germany}

\begin{document}

\section{Introduction}

There are some indications that evolution of AGNs is probably
related with starburst regions (see L\'ipari and Terlevich 2006; Mao et al. 2009; Sani et al.
2010). Namely, it is possible that AGNs in an earlier phase of their evolution are composed
of starburst (star-forming) regions and the central engine, that is AGN, and, during evolution,
the starburst contribution becomes weaker and/or negligible (Wang and Wei 2006, 2008; Mao et al. 2009). On the other hand, it is expected that Fe II strength may 
vary with evolution of AGNs. The optical Fe II ($\lambda\lambda$ 4000-5400 \AA) emission arises from numerous transitions of the complex Fe II ion.
Numerous lines are overlap, forming the complex shapes in $\lambda\lambda$ 4000-5400 \AA \ range and making one of the most interesting features in AGN spectra.  It is seen in almost all type-1 
AGN and it is especially strong in narrow-line Seyfert 1s (NLS1s). Origin of the optical Fe II lines, the mechanisms of their excitation and location of the Fe II emission region in AGN, 
are still open questions (see Kova\v cevi\'c et al. 2010 and references therein). There are also many correlations between Fe II emission and other AGN properties which need a physical explanation.

L\'ipari and Terlevich 2006 have explained some properties of AGN by an ``evolutive unification model''. In this model, accretion arises 
from the interaction between nuclear starbursts and the supermassive black hole. Thus, young AGN have strong Fe II, BALs, weak radio emission, the NLR is 
compact and faint, and broad lines are relatively narrow. In contrast with this, old AGNs have weak Fe II, no BALs, strong radio emission with extended radio lobes, 
the NLR is extended and bright, and the broad lines have greater velocity widths.

In this paper we analyze how the presence/absence of the starburst contribution to the AGN spectra influences to the correlations between different spectral properties, 
specially to the Fe II emission. To investigate the iron lines, we use our model of iron template, based on the physical properties of iron region.

\section{The sample and analysis}

For this investigation we use the sample of 302 AGNs Type 1, selected with criteria described in Kova\v cevi\' c et al. (2010).

The sample has approximately an uniform redshift distribution in the range of z=0--0.7 and negligible host galaxy contribution.

As it is described in Kova\v cevi\' c et al. (2010), the emission lines within the 4400 -- 5500 \AA \ range are fitted with multiple Gaussians, where each Gaussian represent 
contribution from different emission region. The Balmer lines are fitted with three components: a narrow,
an intermediate and a very broad component (H$\beta$ NLR, ILR and VBLR, respectively). We assumed that all narrow lines originate from the same emission region, and consequently 
they have the same width and shift.
The H$\beta$ broad component is taken as the sum of the H$\beta$ ILR and H$\beta$ VBLR components, and FWHM (Full Width Half Maximum) of H$\beta$ line is measured only for the 
H$\beta$ broad line (see Kova\v cevi\' c et al. 2010). The iron lines within  4400 -- 5500 \AA \ range are fitted with template presented in Section 3.1.

The possible contribution of starbursts in AGN Type 1 spectra has been reported, using the BPT diagnostic (BPT - Baldwin, Philips \& Terlevich 1981) in some  investigations: Popovi\' c et al. (2009) found that in the case 
of NLSy1 galaxy Mrk 493, the narrow-line ratios correspond to starbursts rather than to an AGN origin. Similar result is found in paper of Mao et al. (2009) for the three AGN Type 1,
which are mainly excited from starburst in narrow lines, since they belong to the H II part of BPT diagram.

Since we have complete measurements of line parameters only for the narrow
H$\beta$ and [O III] lines in the whole sample (302 AGNs), we accepted a criteria of
R$=$log([O III]/H$\beta$ NLR) = 0.5 as an indicator of the
predominant starburst emission contribution to the narrow emission lines (for details see Popovi\' c \& Kova\v cevi\' c 2011). We divided our
sample into two subsamples: R $<$ 0.5 (91 AGNs, hereafter starburst dominant)
 and R $>$ 0.5 (210 AGNs, hereafter AGN dominant). Then, we performed correlations between measured spectral properties for two subsamples, in order to check if there are some 
 significant differences in correlation coefficients, which may be signature of different physical properties of emission regions.

\section{Results}
\subsection{Fe II template}

For investigation of the origin of iron lines and for analysis of their correlations with other lines, it is necessary to apply a good template which will fit well iron lines within 
$\lambda\lambda$4000-5500 \AA \ range. Also, precise Fe II template is needed to accurately separate the [O III] and Balmer lines, which overlap with Fe II. But, the construction of 
iron template is very difficult since the iron lines form the features of a complex shape. 

A few empirical and theoretical templates are proposed in literature (Veron-Cetty et al. 2004, Dong et al. 2008, Bruhweiler \& Verner 2008), but they can not explain very well all varieties of the Fe II emission. Namely, in the case of some AGN spectra, specially when the blue bump of Fe II emission (4400-4700 \AA) has higher strength than red one (5100-5600 \AA), these models can not enable 
a good fit (see Appendix B in Kova\v cevi\' c et al. 2010).

\begin{figure}
\includegraphics[width=9.cm]{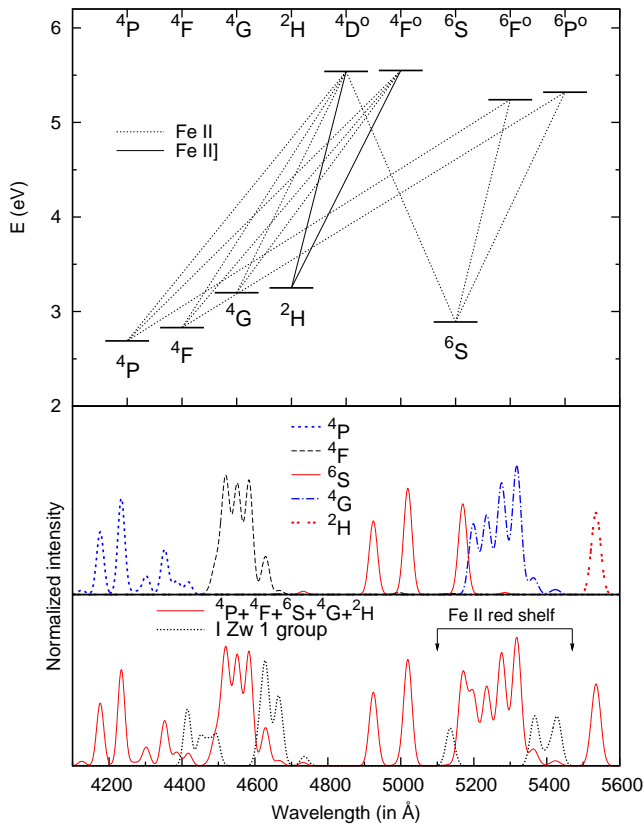}
\caption{The simplified Grotrian diagram showing the strongest Fe II transitions in the 
$\lambda\lambda$ 4100-5600 \AA\ region (top). Lines are separated into five groups according to 
the lower level of transition (middle): P (dotted line), F (dashed line), 
S (solid line), G (dash - dotted line), and H (two-dashed line). Bottom: the lines from the five 
line groups (solid line) and additional lines taken from I Zw 1 (see Kova\v cevi\'c et al. 2010), represented 
with dots. }
\label{G_dijagram}
\end{figure}

We calculated the Fe II template, using 67 Fe II emission lines, identified as the strongest within the $\lambda\lambda$4100-5600 \AA \ range. The 52 of them are
 separated in the five line groups  according to their lower level of transition: $3d^6({\ }^3P2)4s{\ }^4P$, $3d^6({\ }^3F2)4s{\ }^4F$, $3d^54s^2 {\ }^6S $, $3d^6({\ }^3G)4s{\ }^4G$ 
and $3d^6({\ }^3H)4s{\ }^2H$ (in further text ${\ }^4P$, ${\ }^4F$, ${\ }^6S$, ${\ }^4G$ and ${\ }^2H$ group of lines). A simplified scheme of those transitions is shown in Fig \ref{G_dijagram}.

The lines from five line groups describe about 85\% of Fe II emission in the observed range ($\lambda\lambda$4000-5600 \AA), but about 15\% of the Fe II emission can not be explained 
with permitted lines which excitation energies are close to these of lines from the five line groups. The missing parts are around $\sim$4450 \AA, $\sim$4630 \AA, $\sim$5130 \AA \
and $\sim$5370 \AA. 

There are some indications that the process of fluorescence (self-fluorescences, continuum-fluorescences or Ly$\alpha$ and Ly$\beta$ pumping) may have a role in appearing of some 
Fe II lines (Verner et al. 1999, Hartman \& Johansson 2000). They could supply enough energy for exciting the Fe II lines with high energy of excitation, which could be one of the 
explanation for emission in these wavelengths. 

In order to complete the template for missing 15\%, we selected 15 lines, which probably arise with some of these mechanisms, from
Kurutcz database\footnote{http://kurucz.harvard.edu/linelists.html}. The selected lines have wavelengths on missing parts, strong oscillator strength 
and their energy of excitation goes up to $\sim$11 eV. Relative intensities of these 15 lines are addopted from I Zw 1 spectrum by making the best fit 
together with Fe II lines from three line groups. The table of selected lines and their relative intensities are given in Kova\v cevi\'c et al {2010) and Shapovalova et al. (2012).

\begin{figure}
\includegraphics[width=12.cm]{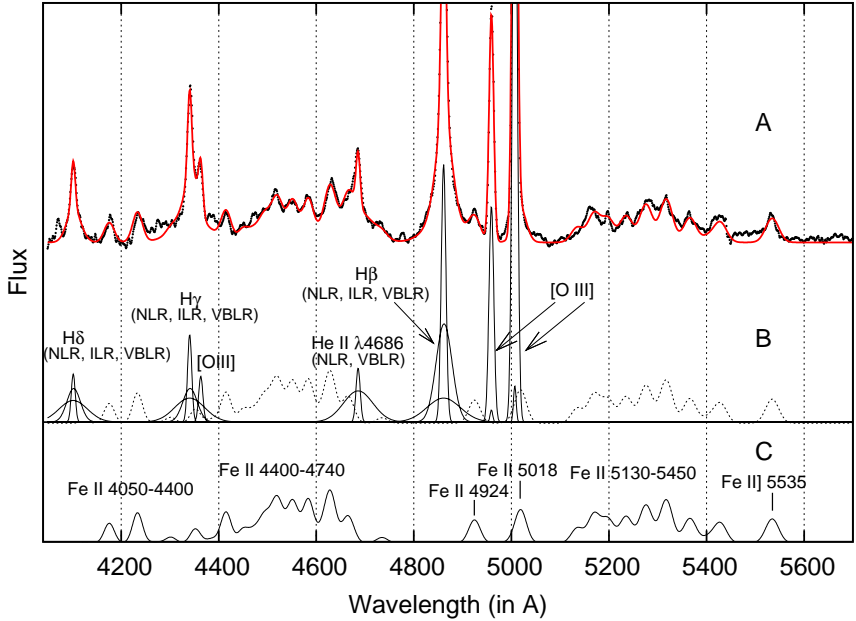}
\caption{An example of the best fit of the $\lambda\lambda$4000-5600 \AA, 
region of Ark 564: (A) the observed spectra (dots) and the best fit (solid line). (B) H$\beta$, H$\gamma$, and H$\delta$ fit 
with the sum of three Gaussians representing emission from the NLR, ILR and BLR. 
The [O III] $\lambda\lambda$4959, 5007 \AA\ lines are fit with two Gaussians for each line of
the doublet and He II $\lambda$4686 \AA\ is fit with one broad and one narrow Gaussian. 
The Fe II template is denoted with a dotted line, and also represented separately in panel (C).}
\label{fig14}
\end{figure}

We have assumed that each of lines can be represented with a Gaussian, described by width (W), shift (d) and intensity (I). Since all Fe II lines from the template 
probably originate in the same region, with the same kinematical properties, values of d and W are the same for all Fe II lines in the case of one AGN, but intensities
are assumed to be different. We suppose that relative intensities between the lines within one line group (${\ }^4P$, ${\ }^4F$, ${\ }^6S$, ${\ }^4G$ or ${\ }^2H$) can be obtained as (Popovi\'c et al.
2009, Kova\v cevi\'c et al. 2010):

\begin{equation}
 \frac{I_1}{I_2}={(\frac{\lambda_2}{\lambda_1})}^3\frac{f_1}{f_2}\cdot\frac{g_1}{g_2}\cdot e^{-(E_1-E_2)/kT}
\end{equation}
where $I_1$ and $I_2$ are intensities of the lines with the same lower level of the transition, $\lambda_1$ and $\lambda_2$ are transition wavelengths, $g_1$ and $g_2$ 
are statistical weights for the upper energy level of the corresponding transition, and $f_1$ and $f_2$ are oscillator strengths, $E_1$ and $E_2$ are energies of the 
upper level of transitions, $k$ is the Boltzman constant and $T$ is the excitation temperature. The list of 52 selected lines from three line groups, their oscillator 
strengths and calculated relative intensities for different excitation temperatures are given in Kova\v cevi\'c et al. (2010) and Shapovalova et al. (2012).

Finally, the template of Fe II is described by 9 free parameters in fit: width, shift, six parameters of intensity - for ${\ }^4P$, ${\ }^4F$, ${\ }^6S$, ${\ }^4G$ and ${\ }^2H$ 
line groups and for lines with relative intensities obtained from I Zw 1. The 9th parameter is excitation temperature included in the calculation of relative intensities within line groups.

The fit of the Fe II template is shown in Fig \ref{fig14}. We apply this template on large sample of AGNs (302) from SDSS database, and we found that the template fits well Fe II emission. 
The Fe II template, as well as web application for fitting Fe II lines in AGN spectra with this model, are given at: http://servo.aob.rs/FeII\_AGN/, as a part of Serbian Virtual Observatory.

\subsection{Influence of starbursts to spectral properties}

We performed correlations between Fe II lines and other spectral properties (continuum luminosity, broad and narrow line widths, equivalent widths of lines) for two subsamples, 
AGN and starburst dominant. We found significant differences in some correlations for these two groups of objects. 

We found that the width of the broad H$\beta$ is in a significant correlation with the  continuum luminosity (L$_{5100}$) for the starburst dominant subsample, while there are
no any correlations between these parameters for the AGN dominant subsample (see Fig 3).

The relation between EW Fe II and FWHM (FWZI) H$\beta$ is also different for the starburst dominant and AGN dominant subsamples (see Fig. 4, Table 1). Namely, the well 
known anticorrelation EW Fe II and FWHM H$\beta$, which is part of EV1 correlations from paper Boroson and Green 1992 (BG92), is noticed only for AGN dominant subsample 
($r_s$=-0.37, P=3.1E-8), while starburst dominant subsample shows even opposite, but statistically not significant trend ($r_s$=0.26, P=0.01). Note here that Grupe (2004) 
also found positive correlation between EW Fe II and FWHM H$\beta$ for NLSy1 objects, and the negative correlation  for BLSy1s. It is the same case for EW [O III] vs. 
EW Fe II anticorrelation, which is also part of BG92 EV1 correlations. Namely it is stronger in AGN dominant subsample ($r_s$=-0.38,  P=8E-9) than in starburst dominant 
($r_s$=-0.27,  P=0.01). Fe II lines show  significant inverse Baldwin effect (see Baldwin 1977) in the AGN dominant subsample ($r_s$= 0.29, P=2.4E-5), while in other subsample ($r_s$=0.26, 
P=0.01) there is no correlation.

As it can be seen in Table 1, there are weak correlations between EW Fe II and the EW of some narrow lines (NLR H$\beta$ and [O III]) only for AGN subsample. Contrary, 
the correlation between EW Fe II and EW broad H$\beta$ is present only for starburst subsample ($r_s$=0.49, P=9.4E-7), while for AGN subsample, no correlation is found 
($r_s$=0.04, P=0.57). 

In principle, the Fe II emission is sligthly stronger (in average) in starburst dominant subsample than in AGN dominant. Average EW Fe II for AGN dominant subsample is: 97 \AA, and for starburst dominant: 119  \AA).

\begin{figure}
\includegraphics[width=0.9\textwidth]{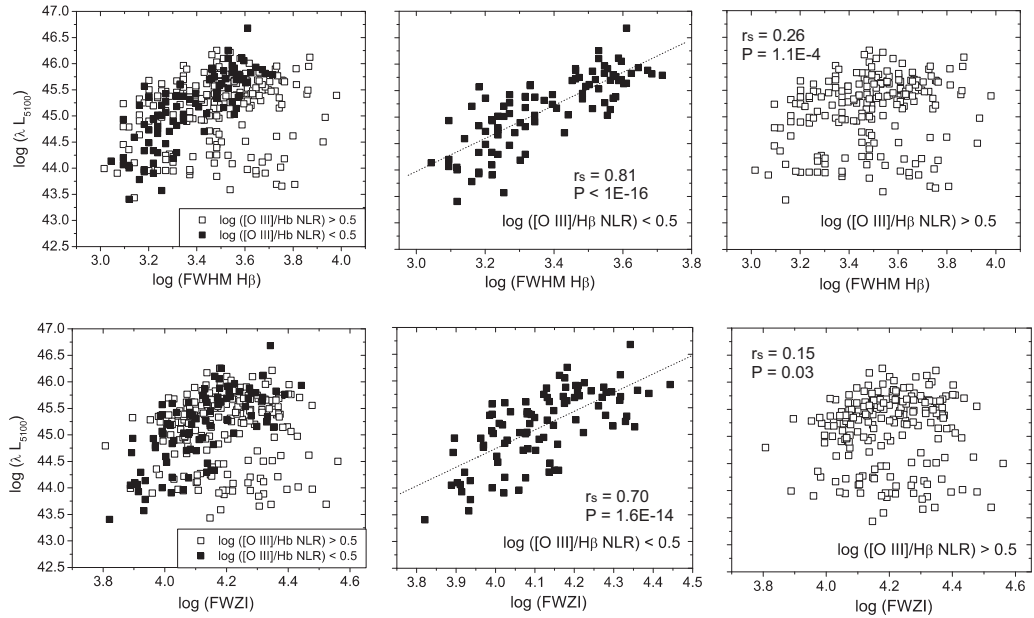}
\caption{Correlations between continuum luminosity and FWHM H$\beta$ (and FWZI H$\beta$) for all AGNs (panels left), AGNs with $R<$0.5 (panels in middle) and $R>$0.5 (panels right).}
\label{f02}
\end{figure}

\begin{figure}
\includegraphics[width=0.9\textwidth]{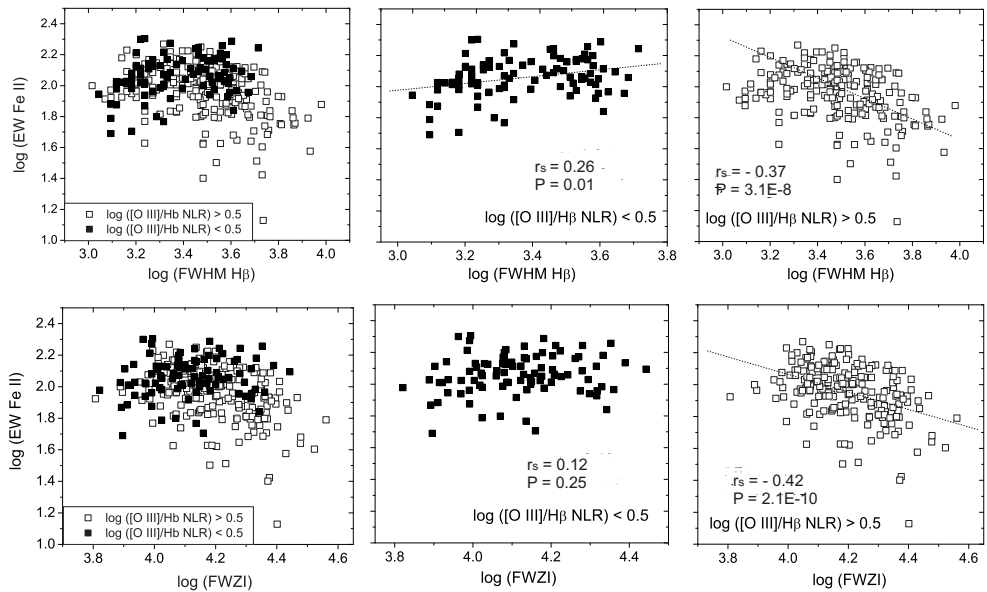}
\caption{The same as in Fig. 3  but for correlations between EW Fe II and FWHM H$\beta$ (and FWZI H$\beta$).}
\label{f03}
\end{figure}

\begin{table}
\begin{center}
\caption{Spearman rank order correlations for  total sample and for subsamples: (1) log([O III]5007/H$\beta$ NLR)$>$ 0.5 (AGN dominant) and (2) log([O III]5007/H$\beta$ NLR)$<$ 0.5 (starburst dominant).}
\begin{tabular}{|c|cc|cc|cc|cc|cc|cc|cc|}
\hline

&&&\multicolumn{2}{|c|}{{\tiny log($\lambda L_{5100}$)}}&\multicolumn{2}{|c|}{{\tiny   log(FWHM H$\beta$)}} & \multicolumn{2}{|c|}{{\tiny  log(EW [O III])}}  &\multicolumn{2}{|c|}{{\tiny   log(EW H$\beta$ NLR)}} &\multicolumn{2}{|c|}{{\tiny  log(EW H$\beta$ broad)}} & \multicolumn{2}{|c|}{{\tiny log(FWMI10\%H$\beta$) }} \\
&& &{\scriptsize r} &{\scriptsize  P }& {\scriptsize r }&{\scriptsize P} &{\scriptsize r }&{\scriptsize P }&{\scriptsize r }&{\scriptsize P }&{\scriptsize r }&{\scriptsize P }&{\scriptsize  r} &{\scriptsize P }\\
\hline

&\multicolumn{2}{|c|}{{\scriptsize    all} }&{\scriptsize  0.27 }&{\scriptsize  1.6E-6 }&{\scriptsize  -0.24 }&{\scriptsize  2.3E-5 }&{\scriptsize  -0.41 }&{\scriptsize  6.8E-14 }& {\scriptsize  -0.01 }&{\scriptsize  0.80 }&{\scriptsize  0.07 }&{\scriptsize  0.23 }&{\scriptsize  -0.26}&{\scriptsize  6.3E-6}\\
 {\tiny log(EW Fe II)   }&\multicolumn{2}{|c|}{{\tiny  (1)} }&{\scriptsize  0.29 }&{\scriptsize  2.4E-5 }&{\scriptsize  -0.37 }&{\scriptsize  3.2E-8 }&{\scriptsize  -0.38 }&{\scriptsize  7.9E-9 }&{\scriptsize  -0.28}&{\scriptsize  3.9E-5 }&{\scriptsize  0.04 }&{\scriptsize  0.57 }&{\scriptsize  -0.38}&{\scriptsize  1.4E-8}\\
 &\multicolumn{2}{|c|}{{\tiny   (2)} } &{\scriptsize  0.26 }&{\scriptsize  0.01 }&{\scriptsize  0.26 }&{\scriptsize  0.01 }&{\scriptsize  -0.27 }&{\scriptsize  0.01 }&{\scriptsize  -0.04 }&{\scriptsize  0.67  }&{\scriptsize  0.49 }&{\scriptsize  9.4E-7 }&{\scriptsize  0.27}&{\scriptsize  0.01}\\
\hline
\end{tabular}
\end{center}
\end{table}

\section{Conclusions}

In this paper we performed correlations between Fe II lines and other spectral properties in the sample of 302 type 1 AGNs. 
We devided the sample in two subsamples, 
AGN dominant and starburst dominant, in order to analyze how the presence/absence of the starburst contribution to the AGN spectra 
influences to the correlations between Fe II and spectral properties.

We found significant difference between all analyzed correlations for different subsamples, which reflex different physics and geometry
of emission region caused by presence/absence of the starbursts. However, there is no big difference in Fe II emission, except that Fe II in starburst dominant subsample tends to be stronger.
The most interesting correlation is between FWHM and luminosity in the subsample of type 1 AGNs where the ratio of narrow lines indicates presence of the star-froming regions.
This may be explained by two possibilities: 1) a non conventional BLR or 2) that AGNs for this subsample are accreating with the similar rate. More detailed discussion can be found in 
Popovi\'c \& Kova\v cevi\'c (2011).

\end{document}